\documentclass[12pt]{article}
\begin{document}
\title{$N=2$ supersymmetric Yang-Mills theory and the superparticle:
twistor transform and $\kappa-$symmetry}
\author{D.V. Uvarov\thanks{E-mail address: uvarov@kipt.kharkov.ua,
d\_uvarov@hotmail.com}\\{\normalsize Kharkov Institute of Physics and Technology}\\ 
{\normalsize 61108 Kharkov, Ukraine}}
\date{}
\maketitle
\begin{abstract}
Lagrangian and Hamiltonian dynamics of de~Azcarraga-Lukierski $N=2$
massive superparticle is considered in the framework of twistor-like
Lorentz-harmonic approach. The emphasis is on the study of the interaction
with external Abelian gauge superfield. The requirement of preservation of
all gauge symmetries of the free model including $\kappa-$symmetry yields correct expressions for the
superfield strength constraints and determines the form of nonminimal
interaction. We also show that for de~Azcarraga-Lukierski $N=2$ massive superparticle the pullback of field strength 2-superform to the superworld line is not integrable in contrast to the massless superparticle.
\end{abstract}

\section{Introduction}

Supersymmetric Yang-Mills theories find the wide range of applications
from a search for possible extentions of the Standard Model to string
theory. The peculiar feature of superfield formulations of supersymmetric
Yang-Mills theories (as well as supergravity theories) is the necessity of the constraints
imposition on superfield strengths (supertorsion
2-form) to eliminate numerous auxiliary fields and achieve agreement with
the component formulations \cite{Wess}, \cite{West}. The justification of
the choice of correct constraints is a subtle matter. It turns out that
the investigation of the interaction of supersymmetric particle models
\cite{BS}, \cite{DAL} can yield proper
superfield constraints for super Yang-Mills theories \cite{Lusanna}, \cite{Rocek}, \cite{Shapiro}, \cite{Delduc}, 
\cite{BN} because the minimal interaction is introduced as the 
pullback of the superpotential 1-form onto the particle's world line.

Another interesting feature of massless superparticles discovered by
Witten \cite{Witten}, \cite{Shapiro} is that they provide the necessary
framework for the twistor transform of the supersymmetric Yang-Mills theories. This transform is based on the integrability of the superfield strength 2-form pullbacked on the light-like superparticle's trajectory in superspace.
For comparison the twistor transform of selfdual Yang-Mills theory
involves 2-complex dimensional null planes \cite{YM}. So, one needs to have
a proper description of supersymmetric light-like lines that possess one
bosonic and $n$ fermionic dimensions and are the trajectories of massless
superparticles in the target superspace. The Grassmann dimensionality of
supersymmetric light-like lines equals to half of that for the target superspace and is related to the partial breaking of global supersymmetry and local $\kappa-$symmetry.

It is explained by the invariance of the massless Brink-Schwarz superparticle \cite{BS} with an arbitrary number $N$ of supersymmetries, as well as, the massive de~Azcarraga-Lukierski superparticle \cite{DAL} with extended $N>1$
supersymmetry under the local $\kappa-$symmetry transformations
\cite{DAL}, \cite{Siegel} which allow to gauge away half of the Grassmann coordinates
$\theta$. In the original formulation this symmetry is infinitely reducible
and the way to remedy this drawback is to introduce auxiliary Lorentz-harmonic variables \cite{VZ}, \cite{Sokatchev}, 
\cite{NPS}, \cite{Bandos'90}, \cite{harmonics}, \cite{BZnull}, \cite{BZstring}, \cite{BZbrane} 
that generalize those
harmonic variables advanced in \cite{GIKOS} to describe theories with
extended supersymmetry in superspace. Lorentz harmonic approach is in fact
the component version of the more general superembedding approach (for
review see e.g. \cite{Sorokin}) that treats branes as supersurfaces
embedded into a target superspace and traces back to the Lund-Regge-Omnes
geometric approach to string theory \cite{LRO}, \cite{Z81}.

Here we develop the Lorentz-harmonic formulation for massive $N\!=\!2$
de~Azcarraga-Lukierski superparticle\footnote{Superparticles based on de~Azcarraga-Lukierski model \cite{DAL} but with the harmonics \cite{GIKOS} related to the automorphisms group of extended supersymmetry rather than the Lorentz group were considered in \cite{ABS}.} in the super Yang-Mills background.
This allows to realize $\kappa-$sym\-met\-ry in the irreducible form, where
its world-line nature is more transparent. The condition for the model
to preserve the $\kappa-$symmetry after the transition to the super Yang-Mills
background requires introduction of nonminimal interaction terms to yield
the desired constraints on the superfield strengths. This nonminimal
interaction amounts to taking into account the superparticle's anomalous
magnetic moment with the value equal to
$\mu=\frac{e}{2m}$ which is fixed by the $\kappa-$symmetry invariance.
In \cite{ZU98} these results were obtained in the framework of the Hamiltonian approach. 

We start from the Lagrangian \cite{ZU98} and analyse the Noether identities applying the above-mentioned Lorentz harmonic technique.
The correspondence between the triviality of superfield strength on the
superworld line and the $\kappa-$symmetry preservation in the presence
of interaction is also examined. 

\section{Lagrangian formulation}
de~Azcarraga-Lukierski massive superparticle in $D=4$ $N=2$ target superspace is described by the action 
\begin{equation}\label{AL}
{\displaystyle
S=-m\int\! \sqrt{-\omega^{\mu}_\tau\omega_{\tau\mu}}
+im\int\! \left(\theta ^{\alpha }_I\dot\theta ^I_{\alpha }
-\bar{\theta}_{\dot\alpha  I}\dot{\bar\theta}\mathstrut^{\dot\alpha  I}\right),}
\footnote{Note that the Wess-Zumino term can also be taken of the form
$m\int(\theta^\alpha_I\dot\theta_\alpha^I+\theta_{\dot\alpha
I}\dot\theta^{\dot\alpha I})$. We use another expression since the factor
$i$ that supersymmetric Cartan forms contain then drops out from the
equations of motion for $\theta^\alpha_I, \theta_{\dot\alpha I}$.}
\end{equation}
where $\omega^m_\tau=\dot x^m+i\theta^\alpha_I\sigma^m_{\alpha\dot\beta}\dot\theta^{\dot\beta I}-i\dot\theta^\alpha_I\sigma^m_{\alpha\dot\beta}\theta^{\dot\beta I}$,
$\dot\theta^I_\alpha$, $\dot\theta^{\dot\alpha I}$ are the pullbacks
onto the world line, parametrized by $\tau$, of the $D=4$ $N=2$
supersymmetric Cartan forms. The automorphisms group of $N=2$ superalgebra
is chosen to be $SU(2)$ and $\theta^I_\alpha$, $\theta^{\dot\alpha I}$
belong to its fundamental representation. The second term in
(\ref{AL}) is the $1d$ Wess-Zumino term ensuring the local
$\kappa-$symmetry invariance of the action. In the first order formalism action (\ref{AL}) is presented as 
\begin{equation}\label{fof}
S=\int d\tau\left[
p_m\omega^m_\tau-\frac12 v(p^2+m^2)\right]+im\int d\tau(\theta^\alpha_I\dot\theta_\alpha^I-\theta_{\dot\alpha I}\dot\theta^{\dot\alpha I}), 
\end{equation}
where $p_m$ is particle's momentum and $v(\tau)$ the Lagrange multiplier imposing the mass-shell constraint $p^2+m^2=0$. 

In order to solve it in the manifestly covariant way and gain irreducible description of the symmetries of superparticle action we introduce appropriate Lorentz-harmonic variables $V=(v_\alpha^{(\mu)}, v_{\dot\alpha}^{(\dot\mu )})\in SL(2,C)$ defined by the harmonicity conditions
\begin{equation}\label{harm}
\Xi=\frac12 v^\alpha_{(\mu)}v_\alpha^{(\mu)}-1=0,\quad \bar\Xi=\frac12
v^{\dot\alpha}_{(\dot\mu)}v_{\dot\alpha}^{(\dot\mu)}-1=0
\end{equation}
which results in $det v_\alpha^{(\mu)}=det
v_{\dot\alpha}^{(\dot\mu)}=1$ .
The inverse harmonics can be expressed as
$V^{-1}=(\epsilon^{\alpha\beta}\epsilon_{(\mu)(\nu)}v_\beta^{(\nu)},
\epsilon^{\dot\alpha\dot\beta}\epsilon_{(\dot\mu)(\dot\nu)}v_{\dot\beta}^{(\dot\nu)})\in
SL(2,C)$, where $\epsilon^{\alpha\beta}, \epsilon^{\dot\alpha\dot\beta}$
and $\epsilon_{(\mu)(\nu)}, \epsilon_{(\dot\mu)(\dot\nu)}$ are $SL(2,C)$
invariant antisymmetric unit tensors. Vector Lorentz harmonics are defined
as bilinear combinations of the spinor Lorentz harmonics
$u^{(n)}_m=\frac12
v_{\dot\alpha}^{(\dot\mu)}\tilde\sigma_m^{\dot\alpha\beta}v_\beta^{(\nu)}\sigma^{(n)}_{(\nu)(\dot\mu)}=\frac12
v^\alpha_{(\mu)}\sigma_{m\alpha\dot\beta}v^{\dot\beta}_{(\dot\nu)}\tilde\sigma^{(n)(\dot\nu)(\mu)}$
and are orthonormal
$u^{(n)}_m u^m_{(k)}=\delta^{(n)}_{(k)}$ as the consequence of the
harmonicity conditions (\ref{harm}).

In the presence of massive particle $SL(2,C)_R$ group acting on indices in
brackets of harmonics is isomorphic to $SO(3)\simeq$ $SU(2)$ so the
spinor harmonics can be presented as $v_\alpha^{(\mu)}=v_\alpha^i$,
$v_{\dot\alpha}^{(\dot\mu)}=-v_{\dot\alpha i}$,
$v^\alpha_{(\mu)}=v^\alpha_i$, $v^{\dot\alpha}_{(\dot\mu)}=v^{\dot\alpha
i}$, where index $i$ corresponds to the fundamental representation of
$SU(2)$. They parametrize the coset space $SL(2,C)/SO(3)\simeq
SL(2,C)/SU(2)$. Accordingly vector harmonics can be divided into
tangential
\begin{equation}\label{2}
u_m^{(0)}=\frac12 v_{\dot\alpha i}\tilde\sigma_m^{\dot\alpha\beta}v^i_\beta=-\frac12 v^\beta_i
\sigma_{m\beta\dot\alpha}v^{\dot\alpha i},\quad u^{(0)}\cdot u^{(0)}=-1
\end{equation}
and orthogonal
\begin{equation}\label{3}
u_m^{(i)}=-\frac12 v_{\dot\alpha i}\tilde\sigma_m^{\dot\alpha\beta}v^j_\alpha\tau^{(i)}{}_j{}^i=-\frac12
v^\beta_i \sigma_{m\beta\dot\alpha}v^{\dot\alpha j}\tau^{(i)}{}_j{}^i,\quad
u^{(i)}\cdot u^{(j)}=\delta^{(i)(j)}
\end{equation}
to the particle's trajectory sets. In (\ref{2}) and (\ref{3}) for brevity
summation over Minkowski vector indices was denoted by $\cdot$ and
$\tau$-matrices were introduced:
$\tau^{(i)}{}_i{}^j=\sigma^{(i)}_{\alpha\dot\beta}=-\tilde\sigma^{(i)\dot\beta\alpha}$.
Indices of the ${\bf3}-$representation of $SU(2)$ are denoted by small
Latin letters in brackets to distinguish from the fundamental
representation indices.

The above analysis suggests that the mass-shell constraint $p^2+m^2=0$ can be solved by putting
$p_m=mu_m^{(0)}$. This relation is the key to the construction of harmonic
formulation for the massive superparticle. The action (\ref{fof}) then reads
\begin{equation}\label{4}
S=m\int
u_m^{(0)}\omega^m_\tau+im\int(\theta^\alpha_I\dot\theta_\alpha^I-\theta_{\dot\alpha I}\dot\theta^{\dot\alpha I}), 
\end{equation}
Analogous Lorentz-harmonic formulation for $D0-$brane in $D=10$ $N=2A$ superspace was considered in
\cite{Bandos'99} using the generalized action approach \cite{BSV}.

The general variation of the action (\ref{4}) is
\begin{equation}\label{5} 
\delta S=-m\int \dot u_m^{(0)}\omega^m(\delta)+m\int\delta
u_m^{(0)}\omega^m_\tau+2im\int(v^{\beta}_{j}\dot\theta^I_\beta+
v^{\dot\beta}_j\dot\theta^I_{\dot\beta})\delta\theta^{+j}{}_I,
\end{equation}
where $\delta\theta^{+j}{}_I=v^{\alpha j}\delta\theta_{\alpha I}+v^{\dot\alpha j}\delta\theta_{\dot\alpha I}$,
$(\delta\theta^{+j}{}_I)^\ast=-\delta\theta^+_j{}^I$, and
$\omega^m(\delta)=\delta x^m+i\theta^\alpha_I\sigma^m_{\alpha\dot\beta}\delta\theta^{\dot\beta I}-i\delta\theta^\alpha_I\sigma^m_{\alpha\dot\beta}\theta^{\dot\beta I}$.
Variation for harmonics requires care since they satisfy orthonormality
conditions (\ref{harm}). The most general variation that does not violate
them is given by
\begin{equation} \delta
u^{(n)}_m=\Omega^{(n)}{}_{(k)}(\delta)u_m^{(k)}:\quad\delta
u^{(0)}_m=\Omega^{(0)(i)}(\delta)u_m^{(i)},\quad \delta
u^{(i)}_m=\Omega^{(0)(i)}(\delta)u_m^{(0)}+\Omega^{(i)(j)}(\delta)u_m^{(j)}.
\end{equation}
For detailed discussion see \cite{BZnull}-\cite{BZbrane}. Expression
(\ref{5}) does not enter the counterpart of $\delta\theta^{+j}_I$:
$\delta\theta^{-j}{}_I=v^{\alpha j}\delta\theta_{\alpha I}-v^{\dot\alpha j}\delta\theta_{\dot\alpha I},$
$(\delta\theta^{-j}{}_I)^\ast=\delta\theta^-_j{}^I$ so that $\frac{\delta
S}{\delta\theta^{-j}_I}\equiv0$. This indicates the invariance of the
action (\ref{4}) under $\kappa-$symmetry transformations in their
irreducible version 
\begin{equation}
\omega^m(\delta)=0,\quad \delta u_m^{(0)}=0,\quad\delta\theta^+_j{}^I=0\Rightarrow
\delta\theta^I_\alpha=-\kappa_i^Iv_\alpha^i,\
\delta\theta^I_{\dot\alpha}=\kappa_i^Iv^i_{\dot\alpha}
\end{equation}
with local parameters $(\kappa_i^I)^\ast=\kappa^i_I$. Other Noether
identities $u^{(0)}_m\frac{\delta S}{\delta\omega_m}\equiv0$,
$\frac{\delta S}{\delta\Omega^{(i)(j)}}\equiv0$ correspond to
reparametrization invariance and local $SU(2)-$symmetry respectively.

To introduce minimal interaction with external Abelian $N=2$ superpotential $A_A(x,\theta,\bar\theta)$ in (\ref{fof}) one
has to shift the coefficients at the pullbacks of Cartan forms by the corresponding components of superpotential 
\begin{equation}\label{shift}
\begin{array}{c}
p_m\rightarrow p_m+ieA_m(x,\theta,\bar\theta),\\
m\theta^\alpha_i\rightarrow m\theta^\alpha_i-eA^\alpha_i(x,\theta,\bar\theta),\\
m\theta_{\dot\alpha i}\rightarrow m\theta_{\dot\alpha i}+eA_{\dot\alpha i}(x,\theta,\bar\theta).
\end{array}
\end{equation}
This yields the following action 
\begin{equation}\label{min}
\begin{array}{rl}
S&={\displaystyle \int d\tau\left[
p_m\omega^m_\tau-\frac12 v(p^2+m^2)\right]+im\int d\tau(\theta^\alpha_I\dot\theta_\alpha^I-\theta_{\dot\alpha I}\dot\theta^{\dot\alpha I})}\\[0.3cm]
&{\displaystyle +ie\int(\omega^m_\tau
A_m(x,\theta,\bar\theta)+\dot\theta^\alpha_IA_\alpha^I(x,\theta,\bar\theta)+\dot\theta^{\dot\alpha I}A_{\dot\alpha I}(x,\theta,\bar\theta))}. 
\end{array}
\end{equation}
Note that in (\ref{shift}) we shifted Grassmann coordinates $(\theta^\alpha_I, \theta_{\dot\alpha I})$ rather than conjugate momenta $(\pi_\alpha^I, \pi^{\dot\alpha I})$ that can be expressed (\ref{25}) in terms of the coordinates $(\theta^\alpha_I, \theta_{\dot\alpha I})$ and $p_m$ and hence are not independent phase-space variables. However, as was argued in \cite{ZU98} such action is not $\kappa-$invariant.

The way to restore $\kappa-$symmetry is to introduce nonminimal interaction by rescaling the mass entering the mass-shell constraint 
\begin{equation}\label{shift+}
m\rightarrow mF,
\end{equation}
where $F(x,\theta,\bar\theta)$ is a gauge-invariant function of the superfield strengths \cite{ZU98}. The action (\ref{min}) then acquires the form 
\begin{equation}\label{nonmin}
\begin{array}{rl}
S&={\displaystyle \int d\tau\left[
p_m\omega^m_\tau-\frac12 v(p^2+(mF)^2)\right]+im\int d\tau(\theta^\alpha_I\dot\theta_\alpha^I-\theta_{\dot\alpha I}\dot\theta^{\dot\alpha I})}\\[0.3cm]
&{\displaystyle +ie\int(\omega^m_\tau
A_m(x,\theta,\bar\theta)+\dot\theta^\alpha_IA_\alpha^I(x,\theta,\bar\theta)+\dot\theta^{\dot\alpha I}A_{\dot\alpha I}(x,\theta,\bar\theta))}. 
\end{array}
\end{equation}

The solution to the generalized mass-shell constraint $p^2+(mF)^2=0$ is obtained in the same way as for the free model $p_m=mFu^{(0)}_m$ and can be viewed as the rescaling of the vector harmonic $u^{(0)}_m$. Accordingly taking into account the relation (\ref{2}) between vector and spinor harmonics it is natural to assume that $F=k\bar k$, where the complex conjugate factors $k(x, \theta, \bar\theta)$ and $\bar k(x, \theta, \bar\theta)$ tending to unity when $e\to0$ rescale spinor harmonics $v^i_\alpha$ and $v_{\dot\alpha i}$.

So, the starting point of our analysis is the following action
\begin{equation}\label{7}
S=m\int F
u_m^{(0)}\omega^m_\tau+im\int(\theta^\alpha_I\dot\theta_\alpha^I-\theta_{\dot\alpha I}\dot\theta^{\dot\alpha I})+ie\int(\omega^m_\tau
A_m+\dot\theta^\alpha_IA_\alpha^I+\dot\theta^{\dot\alpha I}A_{\dot\alpha I}).
\end{equation}
The variation of the action (\ref{7}) is
\begin{equation}\label{8}
\begin{array}{rl} \delta S=&\!\!-{\displaystyle\int
[m\frac{d}{d\tau}(Fu_m^{(0)})\omega^m(\delta)-2iFu_m^{(0)}\delta\theta^\alpha_I\sigma^m_{\alpha\dot\beta}
\dot\theta^{\dot\beta I}+2iFu_m^{(0)}\dot\theta^\alpha_I\sigma^m_{\alpha\dot\beta}\delta\theta^{\dot\beta I}]}\\[0.3cm]
&\!\!+m{\displaystyle\int \omega^m_\tau \delta(F u_m^{(0)})-2im\int (\dot\theta^\alpha_I \delta\theta^i_\alpha-\dot\theta_{\dot\alpha I}\delta\theta^{\dot\alpha I})+ie\int E^A_\tau E^B(\delta)F_{BA}},\\
\end{array}
\end{equation}
where $E^A_\tau=(\omega^m_\tau, \dot\theta^\alpha_I, \dot\theta^{\dot\alpha I})$, $E^B(\delta)=(\omega^n(\delta),
\delta\theta^\beta_J, \delta\theta^{\dot\beta J})$ and $F=dA=\frac12E^A E^B F_{BA}$ is the field strength 2-superform. Introducing harmonic
variables through the completeness relations
$v^\alpha_iv^i_\beta=\delta^\alpha_\beta,$
$v^{\dot\alpha}_iv^i_{\dot\beta}=\delta^{\dot\alpha}_{\dot\beta}$ we
express (\ref{8}) via $\delta\theta^+_i{}^I=k v^{\alpha}_i\delta\theta_{\alpha}^{I}+\bar k 
v^{\dot\alpha}_i\delta\theta_{\dot\alpha}^I$ and $\delta\theta^-_i{}^I=k
v^{\alpha}_i\delta\theta_{\alpha}^{I}-\bar k v^{\dot\alpha}_
i\delta\theta_{\dot\alpha}^I$.
Thus, (\ref{8}) acquires the form
\begin{equation}\label{9} 
\begin{array}{rl}
\delta S=&{\displaystyle\int\left[-mF\dot
u_m^{(0)}+ie\omega^n_\tau(F_{mn}+{\textstyle\frac{im}{e}}(u_m^{(0)}\partial_nF-u_n^{(0)}\partial_mF))\right.}\\[0.3cm]
&\left.+ie\dot\theta^\alpha_I(F_{m}{}_{\alpha}^I+\frac{im}{e}u_m^{(0)}D^I_\alpha
F)+ie\dot\theta^{\dot\alpha I}(F_{m\dot\alpha I}+\frac{im}{e}u_m^{(0)}D_{\dot\alpha I}F)\right]\omega^m(\delta)\\[0.2cm]
&-\frac{ie}{2}{\displaystyle\int\left[\omega^m_\tau(k^{-1}v^\alpha_i(F_{m}{}^I_\alpha+{\textstyle\frac{im}{e}}u_m^{(0)}D^I_\alpha
F)-\bar k^{-1}v^{\dot\alpha}_i(F_{m}{}_{\dot\alpha}^I+{\textstyle\frac{im}{e}}u_m^{(0)}D_{\dot\alpha}^I F))\right.}\\[0.3cm]
&+\dot\theta^\alpha_J(k^{-1}v^\beta_i(F_\beta^I{}^J_\alpha+\frac{2m}{e}(1+k\bar
k^{-1}F)\varepsilon_{\alpha\beta}\varepsilon^{IJ})-\bar k^{-1}v^{\dot\beta}_iF^I_{\dot\beta}{}^J_\alpha)\\[0.2cm]
&\left.+\dot\theta^{\dot\alpha J}(k^{-1}v^\beta_iF^I_\beta{}_{\dot\alpha J}-\bar k^{-1}v^{\dot\beta}_i
(F^I_{\dot\beta}{}_{\dot\alpha J}-\frac{2m}{e}(1+k^{-1}\bar kF)\varepsilon_{\dot\alpha\dot\beta}\delta^I_J))\right]
\delta\theta^{+i}{}_I\\[0.2cm]
&-\frac{ie}{2}{\displaystyle\int\left[\omega^m_\tau(k^{-1}v^\alpha_i(F_{m}{}^I_\alpha+{\textstyle\frac{im}{e}}u_m^{(0)}D^I_\alpha
F)+\bar k^{-1}v^{\dot\alpha}_i(F_{m}{}_{\dot\alpha}^I+{\textstyle\frac{im}{e}}u_m^{(0)}D_{\dot\alpha}^I F))\right.}\\[0.3cm]
&+\dot\theta^\alpha_J(k^{-1}v^\beta_i(F_\beta^I{}^J_\alpha+\frac{2m}{e}(1-k\bar k^{-1}F)\varepsilon_{\alpha\beta}\varepsilon^{IJ})+\bar k^{-1}v^{\dot\beta}_iF^I_{\dot\beta}{}^J_\alpha)\\[0.2cm]
&\left.+\dot\theta^{\dot\alpha J}(k^{-1}v^\beta_iF^I_\beta{}_{\dot\alpha J}+\bar k^{-1}v^{\dot\beta}_i
(F^I_{\dot\beta}{}_{\dot\alpha J}-\frac{2m}{e}(1-k^{-1}\bar kF)\varepsilon_{\dot\alpha\dot\beta}\delta^I_J))\right]
\delta\theta^{-i}{}_I\\[0.2cm]
&+m{\displaystyle\int F\omega_\tau\cdot u^{(i)}\Omega^{(0)(i)}(\delta)}.\\
\end{array}
\end{equation}
From the variation (\ref{9}) we derive the equations of motion
\begin{equation}
\frac{\delta S}{\delta\Omega^{(0)(i)}}=
\omega_\tau\cdot u^{(i)}=0,
\end{equation}
\begin{equation}
\begin{array}{rl}
\displaystyle{\frac{\delta S}{\delta\omega^m}}=&\!\!
-mF\dot
u_m^{(0)}+ie\omega^n_\tau(F_{mn}+\frac{im}{e}(u_m^{(0)}\partial_nF-u_n^{(0)}\partial_mF))\\[0.2cm]
&+ie\dot\theta^\alpha_I(F_{m}{}_{\alpha}^I+\frac{im}{e}u_m^{(0)}D^I_\alpha
F)+ie\dot\theta^{\dot\alpha I}(F_{m\dot\alpha I}+\frac{im}{e}u_m^{(0)}D_{\dot\alpha I}F)=0,
\end{array}
\end{equation}
\begin{equation}
\begin{array}{rl}
\displaystyle{\frac{2}{ie}\frac{\delta S}{\delta\theta^{+i}_I}}=&\!\!\omega^m_\tau(k^{-1}v^\alpha_i(F_{m}{}^I_\alpha+
\frac{im}{e}u_m^{(0)}D^I_\alpha F)-\bar k^{-1}v^{\dot\alpha}_i(F_{m}{}_{\dot\alpha}^I+\frac{im}{e}u_m^{(0)}D_{\dot\alpha}^I
F))\\[0.2cm]
&+\dot\theta^\alpha_J(k^{-1}v^\beta_i(F_\beta^I{}^J_\alpha+\frac{2m}{e}(1+k\bar k^{-1}F)\varepsilon_{\alpha\beta}\varepsilon^{IJ})-\bar k^{-1}v^{\dot\beta}_iF^I_{\dot\beta}{}^J_\alpha)\\[0.2cm]
&+\dot\theta^{\dot\alpha J}(k^{-1}v^\beta_iF^I_\beta{}_{\dot\alpha J}-\bar k^{-1}v^{\dot\beta}_i
(F^I_{\dot\beta}{}_{\dot\alpha J}-\frac{2m}{e}(1+k^{-1}\bar kF)\varepsilon_{\dot\alpha\dot\beta}\delta^I_J))=0,
\end{array}
\end{equation}
\begin{equation}\label{15}
\begin{array}{rl}
\displaystyle{\frac{2}{ie}\frac{\delta S}{\delta\theta^{-i}_I}}=&\!\!\omega^m_\tau(k^{-1}v^\alpha_i(F_{m}{}^I_\alpha+\frac{im}{e}u_m^{(0)}D^I_\alpha F)+\bar k^{-1}v^{\dot\alpha}_i(F_{m}{}_{\dot\alpha}^I+\frac{im}{e}u_m^{(0)}D_{\dot\alpha}^I
F))\\[0.2cm]
&+\dot\theta^\alpha_J(k^{-1}v^\beta_i(F_\beta^I{}^J_\alpha+\frac{2m}{e}(1-k\bar k^{-1}F)\varepsilon_{\alpha\beta}\varepsilon^{IJ})+\bar k^{-1}v^{\dot\beta}_iF^I_{\dot\beta}{}^J_\alpha)\\[0.2cm]
&+\dot\theta^{\dot\alpha J}(k^{-1}v^\beta_iF^I_\beta{}_{\dot\alpha J}+\bar k^{-1}v^{\dot\beta}_i
(F^I_{\dot\beta}{}_{\dot\alpha J}-\frac{2m}{e}(1-k^{-1}\bar kF)\varepsilon_{\dot\alpha\dot\beta}\delta^I_J))=0. 
\end{array}
\end{equation}
The Noether identities for the reparametrization symmetry and local
$SU(2)-$invariance have the form
\begin{equation}
\begin{array}{rl}
u^{(0)}_m\frac{\delta S}{\delta\omega_m}+ie(u^{n(i)}F_{nm}u^{m(0)}+\frac{im}{e}u^{m(i)}\partial_m
F)\frac{\delta S}{\delta\Omega^{(0)(i)}}&+\\[0.2cm]
\frac{1}{\omega_\tau\cdot u^{(0)}}[(k\dot\theta^\alpha_I v^j_\alpha+\bar k\dot\theta^{\dot\alpha}_I v_{\dot\alpha}^j)\frac{\delta S}{\delta\theta^{+j}_I}+(k\dot\theta^\alpha_I v^j_\alpha-\bar k\dot\theta^{\dot\alpha}_I v_{\dot\alpha}^j)
\frac{\delta S}{\delta\theta^{-j}_I}]&\equiv0,\\
\end{array}
\end{equation}
\begin{equation}
\frac{\delta S}{\delta\Omega^{(i)(j)}}\equiv0,
\end{equation}
and hold for any values of the background superfield strength and
functions $k$ and $\bar k$.
When the interaction is switched off $\frac{\delta S}{\delta\theta^{-i}_I}\equiv0$ that reflects the presence of
$\kappa-$symmetry of the action (\ref{4}). Thus, for the action (\ref{7})
to be invariant under $\kappa-$symmetry transformations in the presence of
external background we require (\ref{15}) to be satisfied identically for
any values of $x^m$, $\theta^\alpha_I$, $\theta_{\dot\alpha I}$,
$v^\alpha_i$, $v_{\dot\alpha i}$ when the other equations of motion are
taken into account, i.e. we consider equations (\ref{15}) as restrictions
for the background superfield strengths. In this way we obtain from the
last two lines of (\ref{15}) that contain the lowest dimension $[L]^{-1}$
superstrength components
\begin{equation}\label{16}
F_\alpha^I{}^J_\beta-\frac{2m}{e}(1-k^2)\varepsilon_{\alpha\beta}\varepsilon^{IJ}=0,\
F^I_{\dot\alpha}{}^J_{\dot\beta}-\frac{2m}{e}(1-\bar k^2)\varepsilon_{\dot\alpha\dot\beta}\varepsilon^{IJ}=0,\
F^I_{\alpha\dot\beta J}=0.
\end{equation}
To analyse the content of the constraints (\ref{16}) we decompose
spinor-spinor components of superfield strengths on the
$SU(2)-$irreducible parts
\begin{equation}
F_\alpha^I{}^J_\beta=-\varepsilon_{\alpha\beta}\varepsilon^{IJ}
\bar W+\tau^{(I)IJ}F^{(I)}_{\alpha\beta},\
F^I_{\dot\alpha}{}^J_{\dot\beta}=-\varepsilon_{\dot\alpha\dot\beta}\varepsilon^{IJ}W+
\tau^{(I)IJ}F^{(I)}_{\dot\alpha\dot\beta}.
\end{equation}
Substituting these expansions back into (\ref{16}) yields
\begin{equation}\label{20}
1+\frac{eW}{2m}=\bar k^2,\ 1+\frac{e\bar W}{2m}=k^2,\
F^{(I)}_{\alpha\beta}=0,\ F^{(I)}_{\dot\alpha\dot\beta}=0
\end{equation}
so that
\begin{equation}
k=\sqrt{1+\frac{e\bar W}{2m}},\ \bar k=\sqrt{1+\frac{eW}{2m}},\
F=\sqrt{\left(1+\frac{eW}{2m}\right)\left(1+\frac{e\bar W}{2m}\right)}.
\end{equation}
Solution for
the Bianchi identities for superfield strengths satisfying the constraints
$F^{(I)}_{\alpha\beta}=0$, $F^{(I)}_{\dot\alpha\dot\beta}=0$ and
$F^I_{\alpha\dot\beta J}=0$ yields \cite{Sohnius} that the superfields $W$
and $\bar W$ are chiral $D_{\dot\alpha I}W=D^I_\alpha\bar W=0$ and
spinor-vector superfield strength components are of the form
\begin{equation}\label{22}
F_{m}{}^I_\alpha=\frac{i}{4}\sigma_{m\alpha\dot\beta}D^{\dot\beta I}\bar W,\quad F_{m}{}_{\dot\alpha I}=
-\frac{i}{4}D^\beta_I\sigma_{m\beta\dot\alpha}W.
\end{equation}
Thus, the first line of (\ref{15}) turns to 0 identically.
Note that for the minimal interaction $F=1$ there remains additional
algebraic constraint
$\sqrt{\left(1+\frac{eW}{2m}\right)\left(1+\frac{e\bar W}{2m}\right)}=1$
that eliminates physical degrees of freedom of the superfields $W$ and
$\bar W$.

\section{Hamiltonian formulation}
The Noether identites of the Lagrangian description are in one-to-one
correspondence with the first-class constraint in the Hamiltonian picture.
To discover expressions for these constraints we perform the Hamiltonian
analysis of the model (\ref{7}). To this end we introduce canonical
momenta conjugate to coordinates $Q=(x^m, \theta^I_\alpha, \theta_{\dot\alpha I}, v^i_\alpha, v_{\dot\alpha i})$
\begin{equation}
P=\frac{\partial L}{\partial\dot Q}
\end{equation}
and find the set of the primary constraints
\begin{equation}\label{24}
\Phi_m={\cal P}_m-mFu^{(0)}_m\approx0,
\end{equation}
\begin{equation}\label{25}
V^I_\alpha=\pi^I_\alpha+ip_{\alpha\dot\alpha}\theta^{\dot\alpha I}+im\theta^I_\alpha-ieA^I_\alpha\approx0,\quad 
V_{\dot\alpha I}=\pi_{\dot\alpha I}+i\theta^{\alpha}_{I}p_{\alpha\dot\alpha}-im\theta_{\dot\alpha I}-ieA_{\dot\alpha I}\approx0,
\end{equation}
\begin{equation}
P^i_\alpha\approx0,\quad P_{\dot\alpha i}\approx0,
\end{equation}
where generalized momentum is defined as ${\cal P}_m=p_m-ieA_m$. In
harmonic sector there are two extra constraints (\ref{harm}). To exclude
them from the list of constraints one can utilize covariant momenta in
harmonic sector
\begin{equation}
\Pi^{(k)(l)}=v_\alpha^{(\mu)}\sigma^{(k)(l)}{}_{(\mu)}{}^{(\nu)}P^\alpha_{(\nu)}-P^{\dot\alpha}_{(\dot\nu)}\tilde\sigma^{(k)(l)(\dot\nu)}{}_{(\dot\mu)}v_{\dot\alpha}^{(\dot\mu)}\approx0.
\end{equation}
These $6\!\!=\!\! dimSO(1,3)$ constraints commute with harmonicity
conditions and form the $SO(1,3)$ algebra on the Poisson brackets.
Covariant momenta are decomposed on $SU(2)-$co\-va\-ri\-ant parts as follows
\begin{equation}\label{28}
\Pi^{(0)(i)}=-\frac12 v_\alpha^i\tau^{(i)}{}_i{}^j P^\alpha_j-\frac12
P^{\dot\alpha i}\tau^{(i)}{}_i{}^j v_{\dot\alpha j}\approx0,
\end{equation}
\begin{equation}\label{29}
\Pi^{(i)(j)}=-v_\alpha^i\tau^{(i)(j)}{}_i{}^j P^\alpha_j+
P^{\dot\alpha i}\tau^{(i)(j)}{}_i{}^j v_{\dot\alpha j}\approx0.
\end{equation}
As we shall see below constraints (\ref{28}) belong to the second class,
whereas the constraints (\ref{29}) to the first class and generate on the
Poisson brackets $SU(2)-$symmetry transformations.

To evaluate Poisson brackets of the primary constraints we adopt the
following definitions
\begin{equation}
\{p_m,x^n\}=-i\delta_m^n,\
\{\pi^I_\alpha,\theta^\beta_J\}=-i\delta^\beta_\alpha\delta^I_J,\
\{\pi_{\dot\alpha I},\theta^{\dot\beta J}\}=-i\delta^{\dot\beta}_{\dot\alpha}\delta^J_I,
\end{equation}
\begin{equation}
\{P^i_\alpha, v^\beta_j\}=-i\delta^\beta_\alpha\delta^i_j,\
\{P_{\dot\alpha i}, v^{\dot\beta j}\}=\delta^{\dot\beta}_{\dot\alpha}\delta_i^j
\end{equation}
and find
\begin{equation}
\{\Phi_m,\Phi_n\}=-eF_{mn}-im(u_m^{(0)}\partial_n F-u_n^{(0)}\partial_m F),
\end{equation}
\begin{equation}
\{\Phi_m, V^I_\alpha\}=-eF_m{}^I_\alpha-imu_m^{(0)}D^I_\alpha F,\quad
\{\Phi_m, V_{\dot\alpha I}\}=-eF_{m\dot\alpha I}-imu_m^{(0)}D_{\dot\alpha I}F,
\end{equation}
\begin{equation}
\{V^I_\alpha, V^J_\beta\}=2m\varepsilon_{\alpha\beta}\varepsilon^{IJ}\left(1+\frac{e\bar W}{2m}\right),\quad
\{V_{\dot\alpha}^{I},V_{\dot\beta}^{J}\}=2m\varepsilon_{\dot\alpha\dot\beta}\varepsilon^{IJ}\left(1+\frac{eW}{2m}\right),
\end{equation}
\begin{equation}
\{V^I_\alpha, V_{\dot\beta J}\}=2\delta^i_J{\cal P}_{\alpha\dot\beta}.
\end{equation}
In harmonic sector we have
\begin{equation}
\{\Pi^{(i)(j)},\Pi^{(i')(j')}\}=i(\delta^{(j)[(i')}\Pi^{(i)(j')]}-\delta^{(i)[(i')}\Pi^{(j)(j')]}),
\end{equation}
\begin{equation}
\{\Pi^{(0)(i)},\Pi^{(j)(j')}\}=i\delta^{(i)(j)}\Pi^{(0)(j')}-i\delta^{(i)(j')}\Pi^{(0)(j)},\quad
\{\Pi^{(0)(i)},\Pi^{(0)(j)}\}=i\Pi^{(i)(j)},
\end{equation}
\begin{equation}
\{\Pi^{(0)(i)}, \Phi_m\}=imFu^{(i)}_m,\quad \{\Pi^{(i)(j)}, \Phi_m\}=0,
\end{equation}
\begin{equation}
\{\Pi^{(0)(i)}, V^I_\alpha\}=0,\quad \{\Pi^{(i)(j)}, V^I_\alpha\}=0.
\end{equation}
Canonical Hamiltonian is defined by the expression
\begin{equation}
H_0=\dot x^m p_m+\dot\theta^\alpha_I\pi^I_\alpha+\dot\theta^{\dot\alpha
I}\pi_{\dot\alpha
I}+\Omega^{(0)(i)}_\tau\Pi^{(0)(i)}+\Omega^{(i)(j)}_\tau\Pi^{(i)(j)}-L,
\end{equation}
where $\Omega^{(0)(i)}_\tau$ and $\Omega^{(i)(j)}_\tau$ are the pullbacks onto the world line of Cartan forms
conjugate to covariant momenta. Following
the Dirac method we add to $H_0$ a linear combination of the primary
constraints with arbitrary Lagrange multipliers to obtain the total
Hamiltonian
\begin{equation}
H=\lambda^\alpha_I V^I_\alpha+\lambda^{\dot\alpha I}V_{\dot\alpha I}+a^m\Phi_m+\eta^{(i)}\Pi^{(0)(i)}+
\eta^{(i)(j)}\Pi^{(i)(j)}\approx0.
\end{equation}

The next step is the exploration of the conservation conditions for the
set of the primary constraints (\ref{24}), (\ref{25}), (\ref{28}),
(\ref{29}) $\dot f=i\{f,H\}\approx0$. As the outcome the total Hamiltonian
is presented as a linear combination of the first-class constraints with
arbitrary Lagrange multiplies
\begin{equation}
H=\lambda_{iI}V^{iI}+aT+\eta^{(i)(j)}\Pi^{(i)(j)}\approx0,
\end{equation}
where
\begin{equation}
V^{iI}=k^{-1}v^{\alpha i}V^I_\alpha+\bar k^{-1}v^{\dot\alpha i}V^I_{\dot\alpha}+
\frac{1}{F}\Pi^{(0)(i)}\tau^{(i)}{}_j{}^i(v^{\alpha j}D^I_\alpha\bar k-v^{\dot\alpha j}D^I_{\dot\alpha}k)\approx0
\end{equation}
is the $\kappa-$symmetry generator,
\begin{equation}
\begin{array}{rl}
T=&\!\!
u^{m(0)}\Phi_m+\frac{1}{mF}(ieu^{m(i)}F_{mn}u^{n(0)}-mu^{m(i)}\partial_m F)\Pi^{(0)(i)}\\[0.2cm]
+&\!\!\frac{i}{4}(v^{\dot\alpha i}D^I_{\dot\alpha}k-v^{\alpha i}D^I_\alpha\bar k)(k^{-1}v^{\beta}_{i}V_{\beta I}-\bar
k^{-1}v^{\dot\beta}_{i}V_{\dot\beta I}+
\frac{1}{F}\Pi^{(0)(i)}\tau^{(i)}{}_i{}^j(v^{\beta}_{j}D_{\beta I}\bar k+v^{\dot\beta}_{j}D_{\dot\beta I}k))\approx0
\end{array}
\end{equation}
generates reparametrizations and $\Pi^{(i)(j)}$  local
$SU(2)-$transformations.

The set of the second-class constraints can be chosen as
\begin{equation}
D^{iI}=k^{-1}v^{\alpha i}V^I_\alpha-\bar k^{-1}v^{\dot\alpha i}V^I_{\dot\alpha}\approx0,\quad
\Phi^{(i)}=u^{(i)}\cdot{\cal P}\approx0,\quad
\Pi^{(0)(i)}\approx0.
\end{equation}

Local symmetries of the action (\ref{7}) are generated on the Poisson
brackets by the first-class constraints. $\kappa-$Symmetry transformations
have the form
\begin{equation}
\begin{array}{c}
\delta\theta^I_\alpha=k^{-1}\kappa^{iI}v_{\alpha i},\
\delta\theta_{\dot\alpha I}=\bar k^{-1}\kappa_{iI}v^i_{\dot\alpha},\
\omega^m(\delta)=0,\\
\delta
u^{(0)}_m=F^{-1}\kappa^{iI}u_m^{(i)}\tau^{(i)}{}_i{}^j(v^{\alpha}_{j}D_{\alpha I}\bar k-v^{\dot\alpha}_{j}D_{\dot\alpha I}k)
\end{array}
\end{equation}
with four anticommuting local parameters $\kappa^i{}_I(\tau)$,
$(\kappa^i{}_I)^\ast=\kappa_i{}^I$. For reparametrization symmetry we find
the following transformation laws
\begin{equation}
\begin{array}{c}
\delta\theta^I_\alpha=\frac{iea}{16mFk}h_i{}^I v^{\alpha i},\
\delta\theta_{\dot\alpha I}=-\frac{iea}{16mF\bar k}h^i{}_I v_{\dot\alpha i},\ \omega^m(\delta)=-au^{m(0)},\\[0.2cm]
\delta u^{(0)}_m=\frac{a}{mF}\ u^{(i)}_m(mu^{n(i)}\partial_n F-
ie u^{n(i)}F_{np}u^{p(0)}+\frac{iem}{16}h^{iI}\tau^{(i)}{}_i{}^j(v^{\beta}_{j}
D_{\beta I}\bar k+v^{\dot\beta}_{j}D_{\dot\beta I}k)),
\end{array}
\end{equation}
where $h^{iI}\!=\!\bar k v^{\dot\alpha i} D^I_{\dot\alpha}\bar W-k v^{\alpha i}D^I_\alpha W$, with the local parameter 
$a(\tau)$. Finally the
$SU(2)-$symmetry with parameters $b^{(i)(j)}(\tau)$ affects only harmonics
\begin{equation}
\delta v^i_\alpha=v^j_\alpha b^{(i)(j)}\tau^{(i)(j)}{}_j {}^i,\ 
\delta v^i_{\dot\alpha}=v^j_{\dot\alpha}b^{(i)(j)}\tau^{(i)(j)}{}_j{}^i\Rightarrow
\delta u_m^{(0)}=0.
\end{equation}

\section{Integrability of the superfield strength on the particle's
superworld line and $\kappa-$symmetry}

Let us now consider the property of integrability of superfield strength
$2-$form on the particle's superworld line. As is known the triviality of $N=1$ $D=3,4,6,10$
super Yang-Mills field strength pullbacked to the
supersymmetric world line of massless superparticle, is in one-to-one
correspondence with the superfield constraints $F_{\alpha\beta}=0$
\cite{Witten}, \cite{Shapiro}\footnote{For the treatment of supergravity constraints as integrability conditions see \cite{Bergshoeff}.}. This phenomenon is intimately related to
$\kappa-$symmetry invariance of the massless superparticle that is also
preserved in the background of $N=1$ super Yang-Mills theory and
constitutes the basis for the twistor transform. 

In the case of the massive particle its superworld line is
parametrized by the single bosonic coordinate $\tau$ and four fermionic
coordinates $\eta_i{}^I$ $(\eta_i{}^I{}^\ast=\eta^i{}_I)$. 
As a result of the geometric properties of the $(1,4)-$supermanifold induced by the superworld line embedding \cite{Sorokin}, the einbein one-superform $e^{\hat a}=d\zeta^{\hat m} e_{\hat m}{}^{\hat a}(\zeta)=(e^\tau, e^i{}_I)$ can be chosen to be flat and the superconnection one-form $\hat\Omega_i{}^j$ may be identified with the correspondent form from the harmonic sector. 
The analysis of the embedding coditions 
\begin{equation}\label{emb}
\omega\cdot u^{(0)}=-e^\tau,\quad\omega\cdot u^{(i)}=0,
\quad\frac12(v^{\alpha i}d\theta_{\alpha I}-v^{\dot\alpha i}d\theta_{\dot\alpha I})=e^i{}_I
\end{equation}
results in the general decompositions for the target superspace one-forms
\begin{equation}
\omega_m=e^\tau u^{(0)}_m,\quad
d\theta_{\alpha}^I=v_{\alpha}^{i}e_{i}{}^{I},\quad d\theta_{\dot\alpha I}=v_{\dot\alpha i}e^{i}{}_{I}.
\end{equation}
Then one observes that the field strength $2-$superform pullbacked to the
massive particle's superworld line is nonzero
\begin{equation}
F|_{M^{(1|4)}}=\frac{i}{4}e^\tau e^i{}_I(v^\alpha_i D_\alpha^I
W+v^{\dot\alpha}_i D_{\dot\alpha}^I\bar W)+\frac12 e^i{}_I e_i{}^I(W+\bar W)\not=0,
\end{equation}
where we used the superfield constraints (\ref{16}), (\ref{20}),
(\ref{22}). On the contrary it can be verified that the field strength
$2-$superform pullbacked to the superworld line of $N=2$ massless superparticle 
is trivial on the mass shell. However, the action of the massless superparticle coupled to
$N=2$ Abelian superpotential is not $\kappa-$invariant. Indeed, the rank
of the Grassmann constraints' matrix $G_0$ for free $N=2$ massless
superparticle
\begin{equation}
V^I_\alpha=\pi^I_\alpha+ip_{\alpha\dot\alpha}\theta^{\dot\alpha I}\approx0,
\quad V_{\dot\alpha I}=\pi_{\dot\alpha I}+i\theta^{\alpha}_{I}p_{\alpha\dot\alpha}\approx0
\end{equation}
\begin{equation}
G_0=\delta^I_J\left(\begin{array}{cc} 0_2&p_{\alpha\dot\beta}\\
p^{\dot\alpha\beta}&0_2 \end{array}\right)
\end{equation}
is halved because of the mass shell constraint $p^2\approx0$. Whereas in
the $N=2$ super Yang-Mills background this is not the case due to the 
nontrivial contribution from
$F_\alpha^I{}^J_\beta=-\varepsilon_{\alpha\beta}\varepsilon^{IJ}\bar W$
and $F^I_{\dot\alpha}{}^J_{\dot\beta}=-\varepsilon_{\dot\alpha\dot\beta}\varepsilon^{IJ}W$
\begin{equation}
G_{int}=\delta^I_J\left(\begin{array}{cc} \delta_\alpha^\beta e\bar W&{\cal P}_{\alpha\dot\beta}\\ 
{\cal P}^{\dot\alpha\beta}&-\delta^{\dot\alpha}_{\dot\beta}eW
\end{array}\right),
\end{equation}
where ${\cal P}^2\approx0$.

\section{Conclusion}

We have considered de~Azcarraga-Lukierski massive $N=2$ superparticle in the twistor-like Lorentz
harmonic formulation nonminimally coupled to external Abelian Maxwell
supermultiplet both in Lagrangian and Hamiltonian approaches. In the
Lagrangian approach local symmetries of the superparticle are encoded in
the Noether identities. The introduction of harmonic variables allows to
realize these symmetries (in particular, $\kappa-$symmetry) in the
irreducible form. We established that in the presence of the interaction
the requirement of preservation of all the symmetries (Noether identities)
of the free superparticle identifies the proper constraints on the superfield
strengths that single out $N=2$ extended Maxwell supermultiplet through
the solution of the Bianchi identities. It also fixes the form of the
gauge-invariant nonminimal interaction. 

In the Hamiltonian approach we
deduced irreducible first-class constraints that generate on the Poisson
brackets $\kappa-$symmetry, world-line reparametrizations and local
$SU(2)-$transformations. 

In the last section we addressed the issue of an
interplay between the triviality of the Abelian field strength
2-superform, when restricted to the particle's superworld line, and
$\kappa-$symmetry. In case of $N=1$ $D=3,4,6,10$ massless particle the
condition of triviality of the superfield strength on the superworld line,
that can be viewed as the supersymmetric generalization of ordinary
light-like line, is equivalent to conventional constraints on superfield
strengths. It is also well known that the same constraints can be obtained
when considering the massless superparticle in the background of external
superpotential and requiring the action to be $\kappa-$invariant. However,
when dealing with $N=2$ super Yang-Mills theory the situation changes.
Though the superfield strength remains trivial on the superworld line of
the $N=2$ massless superparticle the action is no more $\kappa-$invariant
in the background of $N=2$ superpotential. The possibility to preserve
$\kappa-$symmetry exists for massive superparticle and is based on the
introduction of gauge invariant nonminimal interaction. As was shown in
\cite{ZU98} it actually corresponds to the interaction caused by
superparticle's anomalous magnetic moment of the magnitude
$\mu=\frac{e}{2m}$.
Therefore, in the case of $N=2$ supersymmetry the $\kappa-$symmetry requirement is not compatible with the triviality condition for the Abelian superfield strength 2-form independently whether the superparticle is massless or massive. 

It is also interesting to compare the dilaton dependence of the action (\ref{nonmin}), (\ref{7}) with that for the $D0-$brane in $D=10$ $N=2A$ supergravity background \cite{CGNSW}. There the role of superpotential $A_A(x,\theta,\bar\theta)$ is played by the RR 1-superform minimally coupled to $D0-$brane. It has nonzero limit when the interaction is switched off producing  $1d$ Wess-Zumino term. Its superfield strength components depend on dilaton superfield $\phi(x,\theta)$ and its derivatives to compensate the $\kappa-$variation of the kinetic Dirac-Born-Infeld term containing the exponent of the dilaton superfield as the scaling in analogy with the function $F$ in our case. $F$ is the function of gauge-invariant superfields $W=\frac14F^{\dot\alpha I}{}_{\dot\alpha I}$ and $\bar W=-\frac14F^\alpha_I{}^I_\alpha$ which leading component is the complex dilaton-axion scalar-field $z(x)=\phi(x)+ia(x)$ of the $D=4$ $N=2$ Maxwell supermultiplet. The fact that $D=10$ $N=2A$ supergravity theory is the dimensional reduction of $D=11$ supergravity suggests that the nonminimal action (\ref{nonmin}), (\ref{7}) (or its generalization to include supergravity background) can also be obtained by dimensional reduction of $D=5$ $N=1$ massless superparticle model minimally coupled to a Maxwell superfield or $D=5$ supergravity\footnote{The role of Kaluza-Klein reduction mechanism to generate nonminimal interactions and anomalous magnetic moment for $D=4$ spinning particles was clarified in \cite{Z85}.}.

Another line for further investigation is to extend the above considered analysis of $N=2$ super Yang-Mills theory to the most interesting case of $N=4$ super Yang-Mills theory due to the AdS/CFT-correspondence conjecture \cite{AdS}. In this case it should be noted, however, that the superfield constraints like (\ref{16}), (\ref{20}), (\ref{22}) put the theory on the mass shell.

\section{Ackwonledgements}

The author is thankful to D.P. Sorokin and A.A. Zheltukhin for the helpful discussions and
careful reading of the manuscript.

\end{document}